\documentclass[twocolumn,showpacs,preprintnumbers,amsmath,amssymb]{revtex4}
\usepackage{graphicx}% Include figure files
\usepackage{dcolumn}% Align table columns on decimal point
\usepackage{bm}% bold math
\begin{document}
\preprint{APS/123-QED}
\title{Development of a relaxation calorimeter for temperatures between 0.05 and 4~K}% Force line breaks with \\
\author{M. Brando}
 \altaffiliation[Current address:~]{Max Planck Institute for Chemical Physics of Solids, N\"othnitzer Str. 40, 01187 Dresden, Germany}
 \email{manuel.brando@cpfs.mpg.de}
\affiliation{Department of Physics, Royal Holloway, University of London, Egham TW20 0EX, UK}
%\affiliation{
%Second institution and/or address\\ This line break forced% with \\}%
\date{\today}
\begin{abstract}
A detailed description of an isoperibol calorimeter for temperatures between 0.05 and 4~K is presented. The proposed setup can provide absolute values of the heat capacity $C$ of small samples (typically 1~mg). The extremely simple design of the sample platform, based on a sapphire substrate, and the experimental setup, which makes use only of a lock-in amplifier and a temperature controller, make the construction of such a calorimeter easy and inexpensive. The thermal-relaxation method is employed, which utilizes a permanent thermal link $k$ between the sample platform and the low-temperature bath. The temperature dependence of $k(T)$ is shown for several platforms throughout the entire temperature range: $k(T)/T$ is nearly constant down to 1~K, where it starts to decrease smoothly. The observed behavior is thoroughly explained by considering the thermal resistances of the platform constituents. A comparison between the values of $k(T)/T$ for platforms based on sapphire and on silver is presented where no significant difference has been observed. Each platform can be assembled to have a particular value of $k/T$ at 1~K. Since the sample relaxation time $\tau \sim C/k$, $k(T)$ can be adjusted to $C(T)$ to give a reasonably fast measuring time: Here, it is demonstrated how this calorimeter can be used in so-called single-shot refrigerators ($^{3}$He or demagnetization cryostats), where the time for a single measurement is limited. In addition, it can be used in moderate magnetic fields $B \leq 10$~T, because the platform constituents are weakly field dependent.
\end{abstract}
\maketitle
\section{\label{sec:introduction}Introduction}
It is often necessary to work with small samples of only few milligrams when studying the thermal properties of solid materials at very low temperatures, as they are easier to cool down. The first modern low-temperature semi-adiabatic calorimeter was developed by Eucken and Nernst (1910), but this technique is restricted to temperatures $T > 2$~K and a minimum sample mass of about $100$~mg.\cite{eucken-1909,nernst-1910} For temperatures $T < 2$~K, calorimeters which employ adiabatic techniques, such as pulse or continuous warming methods,\cite{hemminger-1984,cochran-1966,pinel-1972} require heat switches to allow the sample to be cooled to the lowest temperature and to provide adequate thermal isolation during the measurement. Mechanical heat switches are not suitable when working with small samples at temperatures below 1~K because of the large heating effects induced by friction, while superconducting heat switches cannot guarantee adequate thermal isolation above 1~K.  Thus, improvements have been made to refine experimental techniques which utilize a permanent weak thermal link $k$ between the sample and the low-temperature bath (isoperbol calorimeters), eliminating the need for a heat switch, while providing a reasonably short cooling time.\cite{bachmann-1972,schutz-1974,schwall-1975,gmelin-1979,stewart-1983,ogata-1984}\\
With the development of lock-in-amplifiers, such non-adiabatic microcalorimetry techniques allow measuring the heat capacity of bulk crystals of a mass of typically 1~mg. Experimentalists favour two principal methods: the steady-state ac heating method (AC)~\cite{sullivan-1968,eichler-1979,schmiedeshoff-1987} and the thermal-relaxation (TR) method.\cite{bachmann-1972,schutz-1974,schwall-1975,regelsberger-1986} Recently, a semi-adiabatic compensated heat-pulse (CHP) calorimeter has been developed, which requires an even weaker thermal link to the bath, compared to the TR method, and allows high precision data within a short measuring time.\cite{fisher-1995,wilhelm-2004} The AC method has the advantage of providing heat capacity data as a continuous function of temperature, and it is therefore suitable, e.g., for studying phase transitions. Moreover, it can be used to carry out measurements at constant $T$, while continuously varying other external parameters, such as magnetic field and pressure. This technique does not, however, provide absolute values of the sample heat capacity $C$. This has to be determined in a different way. The TR method, on the other hand, provides heat-capacity data of high accuracy as it measures sample relaxation time $\tau \sim C/k$ while maintaining both, bath temperature and external parameters, constant. The only drawback of this method is that it gets inappropriate close to $1^{st}$ order phase transitions. The CHP method is fast, but restricted to measurements with increasing temperature and does, therefore, not permit measurements at constant $T$ while continuously varying other parameters. A TR isoperibol calorimeter has been chosen because of its versatility and the necessity of making precise measurements of the specific heat capacity under the influence of external parameters at constant $T$.\\
To allow measurements in the range $0.05 \leq T \leq 4$~K on a large variety of small samples of insulating or metallic materials (as metal-oxides, 3d-electron metals, heavy-fermions, spin glasses etc.), the sample platform was constructed with a simple design, developed by G. R. Stewart's group,\cite{sievers-1994} which makes it possible to employ the TR method with high precision, while substantially cutting down costs on the necessary construction materials. In this paper, the construction of such simple and rather inexpensive sample platforms, based on a sapphire substrate, is described, along with the measurement setup, which makes use of only a lock-in amplifier and a temperature controller. Although similar platform designs have already been used by other groups, the temperature dependence of their thermal conductance $k(T)$ is only known for $T > 2$~K.\cite{klemens-1962,greene-1972} In section~\ref{t2:s-therm-conductance}, measurements of $k(T)$ vs. $T$ are shown between 0.05 and 4~K for several platforms. These data can be well fitted by a theoretical model based on the thermal resistances of the platform constituents and on simple geometrical considerations. Experiments were carried out with platforms based on a sapphire substrate (insulating) and on silver (metallic) to investigate the effect of different substrate materials on $k(T)$. No significant difference between the two performances can be observed.\\
Since $\tau \sim C/k$, a large sample mass, or a large $C$, could imply a measuring time of the order of many hours. For this reason, the TR technique is usually mounted in dilution refrigerators, where the temperature can be kept constant permanently. It is demonstrated here that it is also possible to use this kind of technique in so-called single-shot refrigerators ($^{3}$He or demagnetization cryostats), where the measuring time is limited, if the platform material is properly chosen.\\

\section{\label{sec:principle-of-operation}Principle of operation}
In the TR method, the sample is placed on a platform which contains a heater and a thermometer. The thermal link between sample and platform is $k_{2}$. The platform is connected through a small heat link $k_{1}$ to a reservoir at temperature $T_{0}$. 
\begin{figure}[!t]
\includegraphics[clip,angle=0,width=0.48\textwidth]{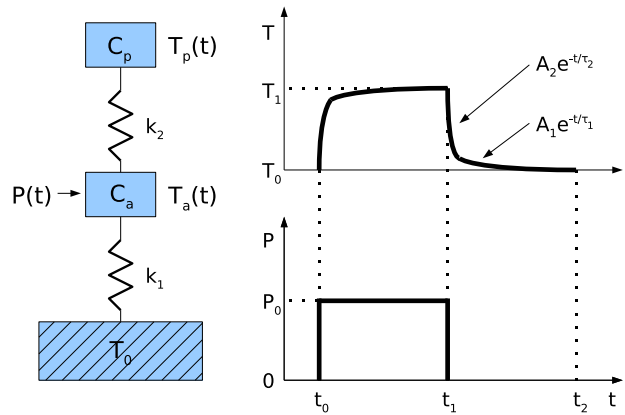}
\caption{(Color online) The RT experiment. \underline{Left}: One-dimensional heat-flow model with a poor thermal contact between sample and platform ($\tau_{2}$ effect): $C_{p}$  and $C_{a}$ are the heat capacities of the sample and platform, respectively. \underline{Right}: Corresponding thermal behavior of the platform thermometer $T_{a}(t)$ during and after the heating pulse.}
\label{fig:model}
\end{figure}
After the bath and the sample have reached a constant temperature $T_{0}$, a well defined constant heating pulse with power $P_{0}$ and duration $t_{1}-t_{0}$ is applied to the platform, until a steady-state temperature $T_{1}$ is reached. The heater power is then turned off and the temperature decays to $T_{0}$ with a time constant $\tau_{1}\approx (C_{p}+C_{a})/k_{1}$. The principle of the experiment is shown in Fig.~\ref{fig:model}, where $C_{p}$ is the heat capacity of the sample and $C_{a}$ is the heat capacity of the platform.\\ 
To analyse this process in detail, it will be assumed assume that the thermal contact between the sample and the platform is good, but not ideal (finite values for $k_{2}$), while the internal thermal conductivity of the platform will be considered ideal. With these assumptions the following one-dimensional heat-flow equations can be solved:
\begin{equation}\label{eq:heat-balance-0}
 \left\{ \begin{array}{l}
P(t)=C_{a}\dot{T}_{a}(t)+k_{1}[T_{a}(t)-T_{0}]+k_{2}[T_{a}(t)-T_{p}(t)]\\
\\
C_{p}\dot{T}_{p}(t)=k_{2}[T_{a}(t)-T_{p}(t)]\end{array} \right.
\end{equation}
where $P(t)$ is the power applied on the platform, $T_{p}(t)$, $T_{a}(t)$ and $T_{0}$ are the sample, platform and bath temperatures; $k_{1}$ and $k_{2}$ are the thermal conductances between platform and bath, and between sample and platform respectively. The heat is flowing in one direction only. If the thermal conductivity of either the sample or the link between sample and substrate is small, compared to that of the link to the bath, the relaxation curves are characterized by a second relaxation time $\tau_{2}$ between sample and platform:\cite{shepherd-1985} These cooling curves show an abnormally high initial slope compared to the rest of the decay (see Fig.~\ref{fig:model} for $t_{1} \leq t \leq t_{2}$). If the sample is a good thermal conductor, but the thermal contact with the substrate is poor, the $\tau_{2}$ effect is called ``lumped''.\cite{shepherd-1985,brando-2000}
\begin{figure}[!t]
\begin{center}
  \includegraphics[clip,angle=270,width=0.48\textwidth]{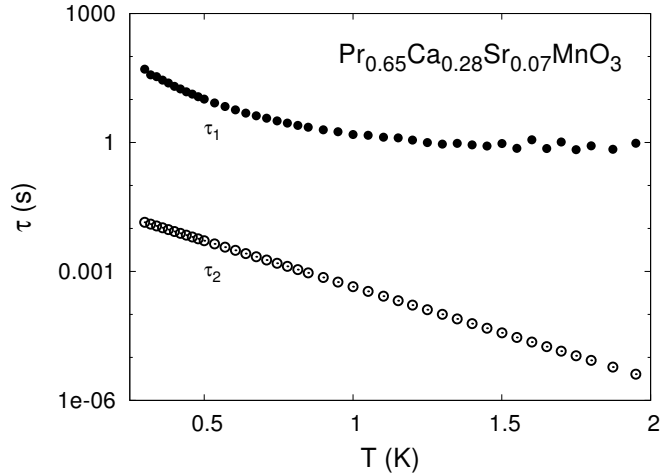}
\end{center}
    \caption{Measurement exhibiting a
    ``lumped'' $\tau_{2}$ effect: The relaxation time $\tau_{2}$ decreases with increasing temperature.\cite{brando-2000}}
\label{t2:f-lumpedtau2}
\end{figure}
\begin{figure}[!t]
\begin{center}
  \includegraphics[clip,angle=270,width=0.48\textwidth]{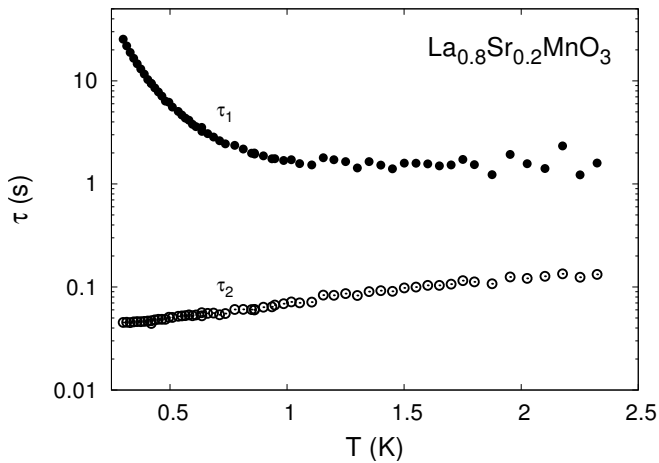}
\end{center}
    \caption{Measurement exhibiting
    a ``distributed'' $\tau_{2}$ effect: The relaxation time $\tau_{2}$ increases with increasing temperature.\cite{brando-2000}}
\label{t2:f-distrtau2}
\end{figure}
In this case the relaxation time $\tau_{2}$ decreases with increasing temperature, as shown in Fig.~\ref{t2:f-lumpedtau2} for a single crystal of Pr$_{0.65}$Ca$_{0.28}$Sr$_{0.07}$MnO$_{3}$.\\
By applying a constant power $P(t)=P_{0}$, and maintaining a maximum temperature rise $\Delta T=T_{1}-T_{0}$ below $2-3\%$, the heat capacity of the sample can be calculated exactly by solving the differential equations~\ref{eq:heat-balance-0} (see Appendix), where the decay in temperature can be represented by a curve consisting of the sum of two exponentials with different time constants $\tau_{1}$ and $\tau_{2}$:
\begin{equation}
 T_{a}(t)=T_{1}-A_{1}e^{-t/\tau_{1}}-A_{2}e^{-t/\tau_{2}}~.\\
\end{equation}
The solution yields:
\begin{equation}\label{eq:solution}
 C_{p}=k_{1}\tau_{1}\left (1-\frac{k_{1}\tau_{2}}{C_{a}}\right ) +k_{1}\tau_{2}-C_{a}~.
\end{equation}
The complete calculation is given in the Appendix.\\
From the experimental value of the heat capacity we have to subtract the heat capacity of the platform $C_{a}$ to obtain $C_{p}$. Since $C_{a}$ can be deduced from preliminary heat-capacity measurements without sample, and since the thermal relaxation time between the sample and the substrate can be calculated from
\begin{equation}\label{eq:tau2}
\tau_{2}=\frac{A_{2}\tau_{1}C_{a}}{(A_{1}+A_{2})\tau_{1}k_{1}-A_{1}C_{a}}~,
\end{equation}
a measure of $A_{1}$, $A_{2}$, $k_{1}$ and $\tau_{1}$ constitutes a measure of $C_{p}$. The thermal conductance $k_{1}$ does not have to be measured every time, but only once, by calibrating the platform, using
\begin{equation}\label{eq:k1}
 k_{1}=P_{0}/(T_{1}-T_{0})=P_{0}/\Delta T~.
\end{equation}
Details of this calibration process are given in Sec.~\ref{t2:s-therm-conductance}.
The parameters $A_{1}$, $A_{2}$ and $\tau_{1}$ can be determined experimentally by analysing the $log(T_{a}(t))$~vs.~$t$ plots of the exponential decay. The main slope of the curve corresponds to $\tau_{1}$, and the initial drop measures the $\tau_{2}$ contribution: With decreasing $\tau_{2}$ effect the drop vanishes.\\
In case the sample has a poor thermal conductivity (as in insulators), different regions of the sample will be at significantly different temperatures. Such a phenomenon is called ``distributed'' $\tau_{2}$ effect.~\cite{shepherd-1985} The solution of the problem is a sum of exponential decays. The time constants of these are given by the solution of a transcendental equation. The calculation of the heat capacity is difficult, but for small effects this contribution can be corrected with the same method used for the ``lumped'' $\tau_{2}$ effect. As shown in Fig.~\ref{t2:f-distrtau2} for a single crystal of La$_{0.8}$Sr$_{0.2}$MnO$_{3}$, the values of $\tau_{2}$ increase with increasing temperature.\cite{brando-2000} If the effect is significant, i.e. if the rate $\tau_{1}/\tau_{2}\rightarrow 1$, this method cannot be used.\\ 
Assuming an ideal thermal contact between sample and substrate ($k_{2}\rightarrow \infty$, $\tau_{2}=0$), the model leads to:
\begin{equation}
\label{eq:solutions}
 \left\{ \begin{array}{l}
T_{a}(t)=T_{1}-A_{1}e^{-t/\tau_{1}}\\
\\
C_{p}=k_{1}\tau_{1}-C_{a}\\
\\
P_{0}=k_{1}(T_{1}-T_{0})=k_{1}A_{1}.\\
\end{array} \right.
\end{equation}
One of the principal advantages of the TR method, e.g. when comparing it with the AC method, is that the corrections for the $\tau_{2}$ effect can be calculated exactly. This is usually valid for $T > 1$, since the ``lumped'' $\tau_{2}$ effect vanishes with increasing temperature (cf.~Fig.~\ref{t2:f-lumpedtau2}). For samples with masses lower than 1~mg, the platform heat capacity $C_{a}$ provides the dominant source of error for $C_{p}$. This explains why some effort must be applied to building platforms with small $C_{a}$ values at low temperature, e.g. using sapphire single crystals as platform substrates.
\section{\label{experimental-setup}Experimental setup}
\begin{figure}[!t]
\begin{center}
 \includegraphics[angle=0,width=0.35\textwidth]{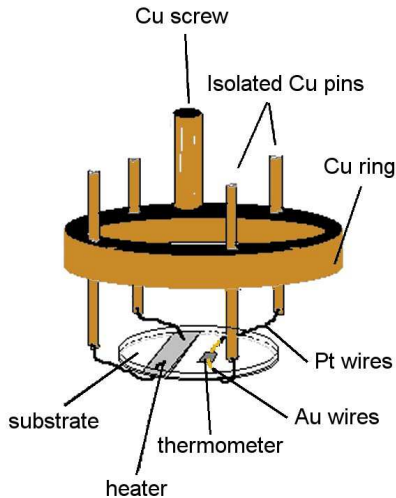}
\end{center}
   \caption{(Color online) General design of the experimental apparatus. The platform is suspended and held mechanically by four Pt wires, which are bonded to electrically isolated Cu-pins inserted in a Cu-ring. The ring is permanently screwed into the low-temperature stage of the cryostat.}
\label{fig:design}
\end{figure}

The apparatus for all heat-capacity experiments is shown in Fig.~\ref{fig:design}: The illustrated design consists of a copper ring-shaped platform holder in which four electrically isolated copper pins are inserted and fixed by a thermally-conducting epoxy cement. The ring is permanently screwed into the low-temperature stage of the cryostat which represents the thermal bath at $T_{0}$. The sample platform consists of a substrate, a heater, a thermometer and bonding silver epoxy.
It is held mechanically in horizontal position by four platinum wires, which are bonded to the copper pins. The wires provide a well defined thermal connection (essentially $k_{1}$) to the isothermal ring, as well as electrical connections for the heater and the temperature sensor attached to the lower side of the platform. In some cases, thin gold wires are used to electrically connect the thermometer with the Pt wires.\\
The data that will be shown hereafter have been collected from measurements carried out with five sample platforms, labelled PLS-1 to 5, mounted in two different systems: a $^{3}$He \emph{Oxford Instruments} Heliox $2^{VL}$ cryostat for $0.3 \leq T \leq 4$~K and a~\emph{Cambridge Magnetic Refrigeration} mFridge for $0.05 \leq T \leq 4$~K.
\subsection{The measurement platforms}
\begin{figure}[!b]
\begin{center}
  \includegraphics[clip,angle=0,width=0.45\textwidth]{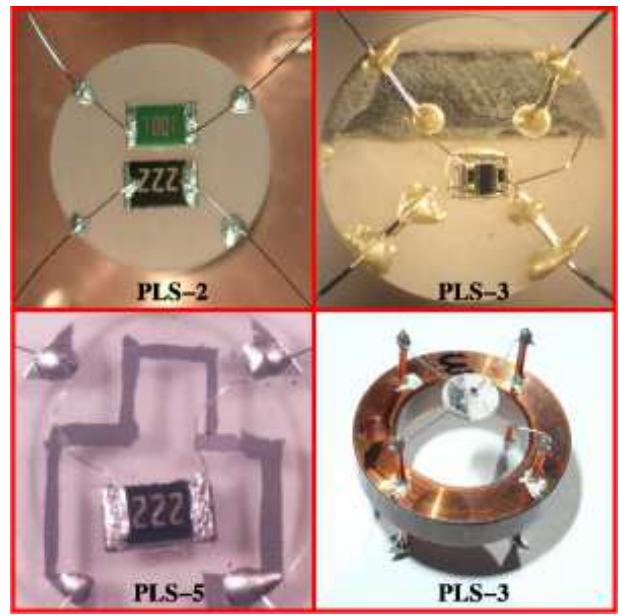}
\end{center}
    \caption{(Color online) Pictures of three of the five sample platforms utilized for this report. The black chips are the RuO$_{2}$ sensors, the green one is the heater chip. On PLS-3, the gray large surface is the sputtered Cr-film heater, which is a squared path on the PLS-5 sapphire disc. On PLS-3 a very small (less than 1~mm$^{3}$) Cernox type sensor is attached to the sapphire surface and connected with tiny gold wires (used also in PLS-5).}
\label{fig:pls0}
\end{figure}

The construction materials are slightly different for each platform; they are listed in Tab.~\ref{tab:materials}. A collection of selected images for three of the platforms are shown in Fig.~\ref{fig:pls0}. The substrate for PLS-1 is a high quality silver (5N) substrate of $100~\mu$m thickness, while for the other platforms sapphire single-crystals of 6~mm diameter and $200~\mu$m thickness were used.\\
\emph{Lake Shore} Cernox type sensors or polished \emph{Bourns} $2$~k$\Omega$ ruthenium-oxide (RuO$_{2}$) resistors were utilized as thermometers (cf. Fig.~\ref{fig:pls0}). The platform temperature $T_{a}$ can be estimated by measuring their resistance. They guarantee high sensitivity below 4~K: Decreasing the temperature down to 0.05~K, the resistance of the RuO$_{2}$ sensors increases from $2$~k$\Omega$ up to values higher than 40~k$\Omega$. In addition, the heat capacity contribution to $C_{a}$ of these thermometers is very small (cf. section \ref{t2:s-addenda-calibration}).\\ 
Polished commercial thin film chips (mass~$\approx 1$~mg, R~$\approx 10$~k$\Omega$) were used as heaters for the first two platforms, while chromium film resistors (with different geometry and resistance of the order of k$\Omega$) were sputtered on one side of the other sapphire discs as heater.
\begin{table}[!t]
\caption{Construction materials.}
\begin{ruledtabular}
\begin{tabular}{lllrr}
Platform & Substrate & Wires & Wire-Diameter & Heater\\
\hline
\\
PLS-1 &  Ag & Pt$_{0.9}$Ir$_{0.1}$ & 50$\mu$m & chip\\
PLS-2 &  Al$_{2}$O$_{3}$ & Pt$_{0.9}$Ir$_{0.1}$ & 50$\mu$m & chip\\
PLS-3 &  Al$_{2}$O$_{3}$ & Pt$_{0.9}$Rh$_{0.1}$ & 50$\mu$m & Cr-film\\
PLS-4 &  Al$_{2}$O$_{3}$ & Pt$_{0.9}$Rh$_{0.1}$ & 125$\mu$m & Cr-film\\
PLS-5 &  Al$_{2}$O$_{3}$ & Pt$_{0.9}$Ir$_{0.1}$ & 50$\mu$m & Cr-film\\
\end{tabular}
\end{ruledtabular}
\label{tab:materials}
\end{table}
The resistance of both kinds of heaters has a very weak temperature dependence and is practically constant below 4~K.\\
The Pt lead wires are made of Ir- or Rh-doped platinum with 50$\mu$m diameter (125$\mu$m for the PLS-4), and provide the well defined thermal link $k_{1}$ to the isothermal ring, as well as electrical connections for the heater and sensor. The Cu-pins in PLS-5 were bent slightly towards the center of the ring to decrease the length of the Pt$_{0.9}$Ir$_{0.1}$ wires, and to thus increase $k_{1}$ (cf.~Fig.~\ref{fig:KTvsT}). By varying the wire length, their thickness or their dopant concentration, the value of $k_{1}$ can obviously be tuned, and, considering $\tau_{1} \sim C_{p}/k_{1}$, be adjusted to what time is available for the experiment.\\
On some of the sapphire substrates, tiny gold wires (25$\mu$m diameter) were spot-welded to the sensor as well as directly to the Pt wires, to reduce $C_{a}$. All constituents were fixed with conducting silver-epoxy cement (\emph{Polytec} Epo-Tex HB1LV). This arrangement ensures an excellent thermal connection between thermometer, heater and substrate. The sample is attached to the back of the sapphire substrate with thermally conductive grease (Apiezon N).\cite{swenson-1999} In the silver platform PLS-1, the wires were directly attached with the same epoxy to the heater and sensor contacts. Soldering was avoided because of the risk of having thin superconducting contacts, which could dramatically reduce the thermal conductance between platform and Cu-ring.
\subsection{The measurement systems}

The measurement system is rather simple and requires just two instruments: A temperature controller to measure and stabilize $T_{0}$ at a thermometer positioned at the bottom of the low-temperature stage and a lock-in amplifier to measure $T_{a}(t)$ and supply the heat pulses.\\
In the Heliox $2^{VL}$ cryostat, $T_{0}$ was measured and controlled with a \textit{RV-Elektroniikka OY} AVS-47 ac resistance bridge and a TS-530 temperature controller. In the mFridge, $T_{0}$ was read by a \textit{Lakeshore} model 340. The direct-temperature-control (DTC) routine, which drives the demagnatisation magnet, seemed the most elegant solution of stabilizing it. In both cases it was possible to stabilize the temperature within 0.2\%.\\
On account of low $\tau_{1}$ and $\tau_{2}$ (e.g. 1~s and 0.05~s, respectively), the voltage signal on the platform sensor has to be read quickly, which is why a lock-in amplifier has been used for measuring the platform thermometer. $T_{a}(t)$ was measured by a \textit{Signal Recovery} 7265 lock-in amplifier connected to the previously calibrated platform sensor through a 10~M$\Omega$ resistance. The lock-in frequency $f$ was set at values close to 100~Hz. It is clear that the lock-in time constant has to be set at a lower value than our $\tau_{1}$ and be comparable to $\tau_{2}$, in order to permit measurement. It can be verified that $\tau_{lock-in}\approx \tau_{1}/40$ provides a good response of the lock-in amplifier. The applied voltage varied from 0.1~V at 50~mK to about 2~V at 3~K to prevent the sensor from self heating and to garantee a good signal reading. The heat pulse was given by the same lock-in amplifier, connecting the DAC output to the platform heater through a 200~k$\Omega$ resistance. High-speed data acquisition was achieved by directly monitoring the output signal with the lock-in buffer option, set at 10~ms. The error due to data noise was successfully reduced to a considerably lower level than the one due to thermal fluctuations.\\
Although the signal offset was large, compared to its changing due to the heat pulse, the quality of the data has been improved by adjusting the lock-in parameters before every single measurement. It was also possible to reduce the signal offset to values close to zero by inserting another RuO$_2$ thermometer on the Cu ring and, with both sensors, building a standard ac Wheatstone bridge, which is driven directly by the lock-in amplifier reference oscillator. Doing that, the resolution of the voltage reading increased by a factor of ten, allowing the bridge to operate now at higher frequency, optimally at $f \approx 1500$~Hz. Values for $\tau_{2}$ lower than 1~ms have thus been detected.\\
As the platform thermometer is calibrated, a reading of $C_{p}$ can always be obtained, along with a measurement of $k_{1}$, with the configuration described above. A major advantages of utilizing the TR method, is that the thermometer on the sample platform is only necessary for measuring relaxation time constants and, therefore, it does not always have to be calibrated. In systems with various measurement platforms assembled together, or with two different thermometers (for separate temperature ranges) on the same platform, this can be a very convenient feature. The thermal conductance $k_{1}$ for each platform has then to be measured separately once (see section \ref{t2:s-therm-conductance}) and the base temperature $T_{0}$ may be detected by the low-temperature stage thermometer of the cryostat.\\

\section{\label{experimental-results}Experimental results}
\subsection{Measurement of the platforms thermal conductance}\label{t2:s-therm-conductance}
\begin{figure}[!b]
\begin{center}
   \includegraphics[angle=0,width=0.48\textwidth]{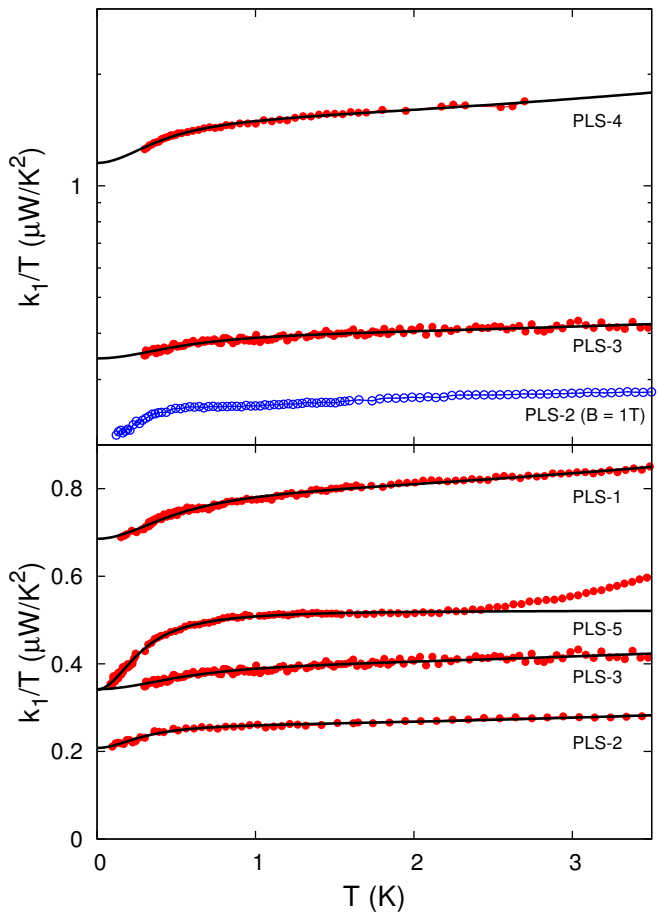}
\end{center}
    \caption{(Color online) Upper frame: Comparison of the thermal conductance for PLS-3 and PLS-4 with two different sizes for the diameter of the lead wires: $50~\mu$m and $125~\mu$m, respectively. The results for PLS-2 in a magnetic field of 1~T are also shown. Lower frame: Measurement of the thermal conductance for platform PLS-1,-2,-3 and PLS-5. The black lines represent the fit curves resulting from equation~\ref{t2:s-thermal-cond-model}. The fit parameters are quoted in Tab.~\ref{tab:parameters}.}
\label{fig:KTvsT}
\end{figure}

The thermal conductance $k_{1}(T)$ was determined by measuring the supplied power $P_{0}$ and the temperature difference $\Delta T = (T_{1}-T_{0})$, as indicated in Eq.~\ref{eq:k1} . $\Delta T$ was kept below 3~\%. As the thermometers of all platforms had previously been calibrated, $k_{1}(T)$ was measured each time, along with the specific heat.\\
With a suitable choice of lead wires, the heat link can be controlled and varied: The upper frame of Fig.~\ref{fig:KTvsT} shows the thermal conductance for the PLS-3 with Pt$_{0.9}$Rh$_{0.1}$ lead wires of $50~\mu$m diameter, and for the PLS-4 with $125~\mu$m of diameter. A relaxation time of these platforms (without a sample) was measured at less than 400~ms even for the lowest temperatures. It is useful to note that, according to Eq.~\ref{eq:solution}, the first platform allows measuring smaller samples, because of its smaller $k_{1}$.\\
Instead of very pure platinum wires, Rh- and Ir-doped ones were chosen, mainly for two reasons: Their robustness and the lower thermal conductivity values.
In fact, the thermal conductivity of a metal at low temperature is strongly influenced by impurities in the material, which tend to decrease $k_{1}$. In addition to that, the thermal conductivity of these wires is not very sensitive to magnetic fields of a relative magnitude ($B\leq 10$~T). This allows measurements of $Cp(T)$ vs. $T$ in magnetic field with the same platforms. A measurement of $k_{1}(T)$ vs. $T$ for the PLS-2 in $B = 1$~T leads to the same results as in zero field. To avoid the overlapping with the data in zero field, shown in the lower frame of Fig.~\ref{fig:KTvsT}, this measurement is plotted in the upper frame of the same figure.\\
The lower frame of Fig.~\ref{fig:KTvsT} shows the thermal conductance for PLS-1,-2, and -5, down to 50~mK and, for PLS-3 down to 300~mK. In all curves, $k_{1}(T)/T$ begins to decrease below 1~K; in PLS-5, this drop is very pronounced. With a simple theoretical model it is possible to explain this effect and to calculate the thermal conductance of the Pt and Au wires, as well as that of the platform substrate. Not only the sapphire disc (described here as substrate) is relevant for this model, but also the insulating substrates of the sensors and all parts where there is a thermal boundary resistance between metal and insulator. We need to consider that the low-temperature thermal conductivity of a metal is primarily due to conduction electrons, and is known to follow a linear-in-$T$ dependence for $T\rightarrow 0$. It is larger than the thermal conductance of an insulator, which is solely due to lattice phonons and follows a $T^{3}$-power law for $T\leq \Theta_{D}$ ($\Theta_{D}$ is the Debye temperature).
\begin{figure}[!t]
\begin{center}
   \includegraphics[angle=0,width=0.45\textwidth]{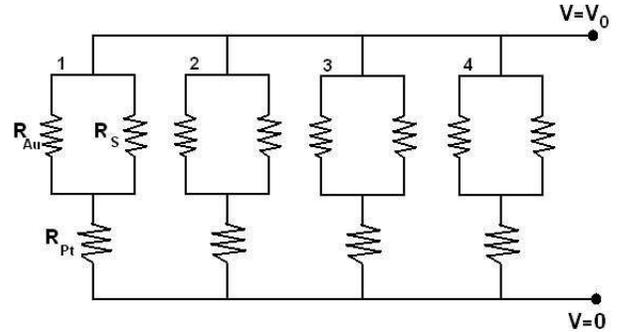}
\end{center}
    \caption{Schematic representation of the thermal resistances on the sample platform: R$_{S}$, R$_{Au}$ and R$_{Pt}$ are the resistances of the substrate, of the gold and platinum wires, respectively.}
\label{t2:k-model}
\end{figure}
If we suppose that the heat originates from the center of the platform and flows simultaneously through the Au wires and the substrate of the platform, and after that through the platinum wires, a simple schematic representation of all thermal resistances can be drawn as in Fig.~\ref{t2:k-model}. The numbers in the picture indicate the four wires which are considered identical. The thermal resistance along every wire is:
\begin{equation}
R_{tot}=R_{Pt}+\frac{R_{S}R_{Au}}{R_{S}+R_{Au}}
\end{equation}
and the relative thermal conductance is:
\begin{equation}
k=\frac{k_{Pt}(k_{S}+k_{Au})}{k_{Pt}+k_{S}+k_{Au}}~.
\end{equation}
Considering the following temperature dependence for the gold and platinum wires and for the substrate contribution:
\begin{equation}
\left\{ \begin{array}{l}
k_{Au}=aT+bT^{3}\\
k_{Pt}=a_{1}T+b_{1}T^{3}\\
k_{S}=b_{2}T^{3}
\end{array}\right.
\end{equation}
the temperature dependence of $K$ can be written as:
\begin{equation}\label{t2:s-thermal-cond-model}
\frac{k(T)}{T}=\frac{aa_{1}+(a_{1}b_{2}+a_{1}b+ab_{1})T^{2}+(b_{1}b_{2}+bb_{1})T^{4}}{(a+a_{1})+(b+b_{1}+b_{2})T^{2}}~.
\end{equation}
Since we have four wires, $k(T)=4k_{wire}(T)$. The derived function fits well the experimental data for PLS-1 to -4 (cf. Fig.~\ref{fig:KTvsT}): The resulting parameters are shown in Tab.~\ref{tab:parameters}.\\ Taking the fit parameters for PLS-5 at 1~K, and supposing that the total length of the four platinum wires is 2~cm, the corresponding thermal conductivity of every wire is about 53~mW/Kcm. For the sapphire thermal conductivity $K_{Sh}$, we can perform a similar calculation, assuming that $b\approx 20~\mu$W/K$^{4}$ at 1~K, and that the length of four conducting paths from the center of the platform to its edges is $l=0.3\times 4$~cm, with a flow area $A=200~\mu$m$\times 1$~mm: We obtain a reasonable value of $K_{Sh}=bl/A=12$~mW/Kcm.\cite{pobell-1996}\\
This model cannot explain the experimental data above 2.4~K for PLS-5. Since the sapphire substrate and the heater design of this platform are different from the others, it is possible that the measured data above 2.4~K are subject to a systematic error. One possibility is that the well polished surface of the sapphire does not allow the epoxy to stick well and creates large thermal boundary resistances. It might also be possible to bond the wires using the technique described in Ref.~\cite{varmazis-1978} to avoid this problem, and to afterwards fix them with epoxy. If we look at the heater shape, on the other hand, we can see that it is closer to the sapphire substrate border than to the thermometer. This design was meant to create a better steady state on the entire platform, but above 2.4~K it may cause the thermometer to be at a lower temperature than the heater. Specific-heat measurements on gold samples also confirm that the data for $k_{1}(T)$ above 2.4~K are not correct.
\begin{table}[t]
\caption{\label{tab:parameters}Fitting parameters.}
\begin{ruledtabular}
\begin{tabular}{cccccccc}
Platform & $a$ & $b$ & $a_1$ & $b_1$ & $b_2$\\
 & ($\mu$W/K$^2$) & ($\mu$W/K$^4$) & ($\mu$W/K$^2$)& ($n$W/K$^4$) & ($\mu$W/K$^4$)\\
\hline
\\
PLS-1 &  4.75 & 0 & 0.80 & 4.19 & 20.24\\
PLS-2 &  1.01 & 0 & 0.26 & 1.69 & 13.30\\
PLS-3 &  2.28 & 0 & 0.40 & 1.87 & 7.75\\
PLS-4 &  4.54 & 0 & 1.54 & 20.36 & 28.80\\
PLS-5 &  0.98 & 0 & 0.52 & 0 & 18.08\\
\end{tabular}
\end{ruledtabular}
\end{table}
Although the reason for the behavior of PLS-5 has not yet been understood, the fact that the silver platform PLS-1 behaves exactly like PLS-2, -3 and -4 indicates that such behavior might derive from the sapphire substrate.\\
Comparing the temperature behavior and the performance of the silver platform with that of the other sapphire platforms, it becomes clear that a sapphire substrate can be used perfectly well instead of a metallic silver one for this kind of heat-capacity measurements (at least down to 50~mK).
\subsection{Calibration}\label{t2:s-addenda-calibration}
The advantage of our platform design is that its contribution $C_{a}$ to the total heat capacity is small. It consists only of the substrate (sapphire was chosen because of its low specific heat at low temperatures), a small amount of grease, tiny polished thermometer and heater chips, and $1/3$ of the lead wires (cf. Ref.\cite{bachmann-1972}).
\begin{figure}[!b]
\begin{center}
   \includegraphics[angle=-90,width=0.48\textwidth]{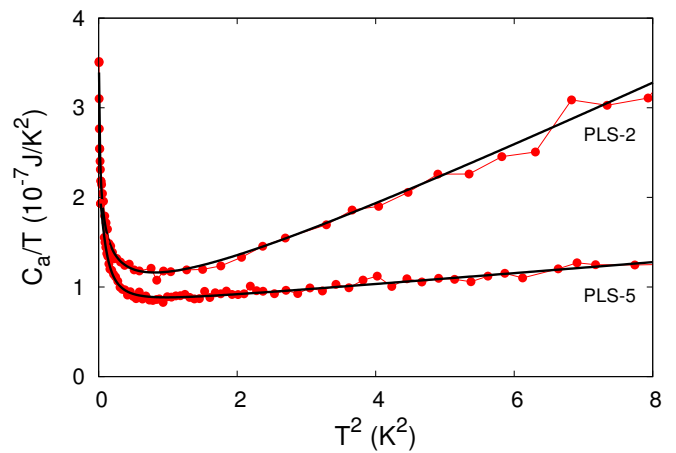}
\end{center}
    \caption{(Color online) Heat capacities of platform PLS-2 and PLS-5. The black lines are fits with the Eq.~\ref{t2:e-addenda-function}.}
\label{t2:f-addenda}
\end{figure}
Accurate values for the specific heat of every single platform constituent can be determined separately or found in literature.~\cite{swenson-1999,pobell-1996} Moreover, it is possible to use very small samples and still keep the platform heat capacity well below the sample's one: e.g. the platform PLS-3 consists of a 34~mg sapphire substrate, a 4~mg Cernox thermometer, 0.678~mg silver-epoxy cement, 0.711~mg gold wires and 6.2~mg Pt$_{0.9}$Rh$_{0.1}$ wires for an amount of heat capacity at 1~K of about $1.7231\times 10^{-7}$~J/K. This value is about 3.5 times lower than the heat capacity of 100~mg gold at 1~K, which is about $5.84\times 10^{-7}$~J/K.\\
Since the specific heat of the Pt$_{0.9}$Ir$_{0.1}$ wires was unknown, two calibration measurements were carried out with two gold samples (purity of 99.99\%) of different weight, 59.1~mg and 72.6~mg. After having subtracted the literature data of Ref.~\cite{martin-1973} for the measured 59.1~mg gold sample, the total heat capacity of the platform is obtained: Its behavior for PLS-2 and PLS-5 can be seen in Fig.~\ref{t2:f-addenda}. For $1.5 \leq T \leq 4$~K, $C_{a}/T$  versus $T^{2}$ is linear, as expected, but below 1.5~K it increases sharply. This effect is very common and due to magnetic impurities in the platform constituents (e.g. in the alumina substrate of the RO$_{2}$ chips) and in the Pt wires.~\cite{pobell-1996,ho-1965} This contribution can be fitted well by adding a $T^{-2}$ Schottky factor to the expected fit function, corrected by a constant $d$, which ideally should be zero:
\begin{equation}\label{t2:e-addenda-function}
\frac{C_{a}}{T}=\gamma+\beta T^{2} + \frac{\delta}{d+T^{3}}~.
\end{equation}
The results of the fits are shown as black lines in Fig.~\ref{t2:f-addenda}. There is a big difference between the heat capacities of the two platforms PLS-2 and PLS-5, as in the PLS-2 a chip has been attached as heater instead of a Cr-film,
\begin{figure}[!t]
\begin{center}
   \includegraphics[angle=-90,width=0.48\textwidth]{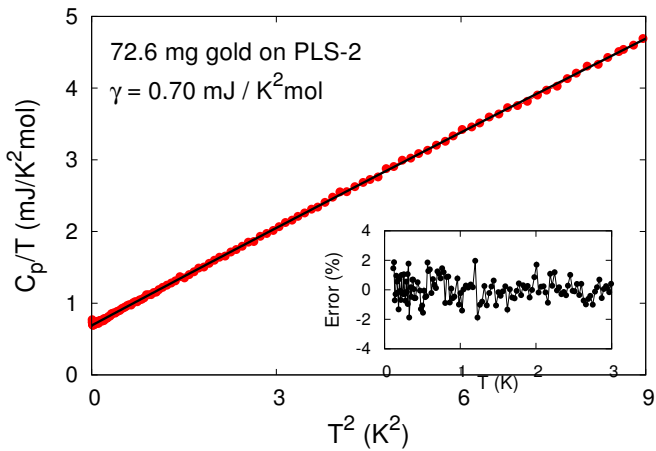}
\end{center}
    \caption{(Color online) Measurement of a 72.6~mg gold sample on PLS-2 after having substracted the platform $C_{a}$. Inset: Relative percent error defined as $(C_{p}-C_{lit.})/C_{lit.}$ where $C_{lit.}$ is the gold heat capacity taken from Ref.~\cite{martin-1973}.}
\label{t2:f-gold}
\end{figure}
and a larger amount of silver epoxy has been used. However, even on PLS-2, the total heat capacity of the platform remains very small when compared to the sample.\\
Fig.~\ref{t2:f-gold} shows the measurement of the second gold sample on PLS-2, after having substracted the platform heat capacity $C_{a}$, in $C_{p}/T$ versus $T^{2}$. The linear fit results in a Sommerfeld coefficient $\gamma = 0.70$~mJ/K$^{2}$mol, very close to the one measured in Ref.~\cite{martin-1973} (0.69~mJ/K$^{2}$mol). The total error is displayed in the inset; it is less than 2\% within the whole temperature range. It is mainly due to the instability of the temperature at all points. Applications of the technique described in this article can be seen, e.g. in Ref.~\cite{knebel-1999,brando-2008}, where samples with masses between 0.2 and 10~mg have been used.
\section{\label{conclusions}Conclusions}
A large number of techniques have been developed to measure the heat capacity of small samples at low temperatures. This paper presented the progress made in improving the TR method. The proposed inexpensive platform design and the simple measurement procedure, which makes use of only two instruments, offer a new prospective for the low-temperature laboratories which suffer of limited funding. The distinct advantage of the method described here, is that it simultaneously provides high-precision measurements, absolute heat capacity values, a certain flexibility in varying temperature or external parameters, while performing the measurements at constant $T$.\\
There are limitations of the TR method, including the fact that the temperature must be kept constant during measurement time, which is why this method has not been used in single-shot refrigerators. In this article it has been demonstrated that this kind of method can be used down to 0.05~K in single-shot refrigerators, too, i.e. $^{3}$He and demagnetization cryostats, provided that materials for the construction of the sample platforms are properly chosen. Detailed information on how to build sample platforms utilizing inexpensive and effective materials has been given, along with the calibration results of the platform heat capacities.\\
Platforms with sapphire and silver substrates were used. For the first time, measurements of thermal conductance $k_{1}(T)$ vs. $T$ for such platforms have been shown below 2~K. A simple theoretical understanding of its behavior has been proposed. Our results indicate that the behavior of the platforms with different substrates are comparable across the entire range of temperature investigated.\\
Finally, measurements of the platform thermal conductance $k_{1}(T)$ were carried out in magnetic field and it could been observed that the magnetic field has no influence on the $k_{1}(T)$ vs. $T$ behavior. This is due to the fact that the thermal conductivity of the lead wires is almost field independent.
\begin{acknowledgments}
I would like to acknowledge F.~M.~Grosche, J.~Hemberger, G.~Knebel, F.~Mayr, M.~Nicklas, E.-W.~Scheidt, W.~Trinkl and R.~Wehn for their precious help and suggestions, C.~Klingner, T.~L\"uhmann, M.~Sugrue and N.~Rothacher for having examined the manuscript.
\end{acknowledgments}
\appendix
\section{Solution of the one-dimentional model}
The thermal equations for the model depicted in Fig.~\ref{fig:model} can be written as:\cite{shepherd-1985}
\begin{equation}
\label{eq:heat-balance}
 \left\{ \begin{array}{l}
P(t)=C_{a}\dot{T}_{a}(t)+k_{1}[T_{a}(t)-T_{0}]+k_{2}[T_{a}(t)-T_{p}(t)]\\
\\
C_{p}\dot{T}_{p}(t)=k_{2}[T_{a}(t)-T_{p}(t)]\end{array} \right.
\end{equation}
Rearranging yields:
\begin{displaymath}
\left \{ \begin{array}{l}
k_{2}T_{p}(t)=-[P(t)+k_{1}T_{0}]+(k_{1}+k_{2})T_{a}(t)+C_{a}\dot{T}_{a}(t)\\
\\
C_{p}\dot{T}_{p}(t)=k_{2}[T_{a}(t)-T_{p}(t)]\end{array} \right.
\end{displaymath}
Considering $P(t)=P_{0}=const$ for $t_{0}\leq t < t_{1}$ and
$P(t)=0$ for $t_{1} \leq t < t_{2}$, we derive the first equation:
\begin{displaymath}
 \left\{ \begin{array}{l}
k_{2}\dot{T}_{p}(t)=(k_{1}+k_{2})\dot{T}_{a}(t)+C_{a}\ddot{T}_{a}(t)\\
\\
C_{p}\dot{T}_{p}(t)=k_{2}T_{a}(t)-k_{2}T_{p}(t)\end{array} \right.
\end{displaymath}
Substituting $T_{p}(t)$ and $\dot{T}_{p}(t)$ into the second equation we obtain:
\begin{equation}\label{eq:main-equation}
 \frac{C_{p}C_{a}}{k_{1}k_{2}}\ddot{T}_{a}(t)+\left [\frac{C_{p}+C_{a}}{k_{1}}+\frac{C_{p}}{k_{2}}\right ]\dot{T}_{a}(t)+T_{a}(t)=T_{0}+\frac{P(t)}{k_{1}}
\end{equation}
If a power $P_{0}$ is applied between $t_{0}$ and $t_{1}$, the platform temperature $T_{a}$ will rise to $T_{1}$ (see Fig.~\ref{fig:model}), according to the following relation:
\begin{equation}
\label{eq:solution-0}
 \left\{ \begin{array}{l}
 T_{a}(t)=T_{1}-A_{1}e^{-t/\tau_{1}}-A_{2}e^{-t/\tau_{2}}\\
\\
\dot{T}_{a}(t)=\frac{A_{1}}{\tau_{1}}e^{-t/\tau_{1}}+\frac{A_{2}}{\tau_{2}}e^{-t/\tau_{2}}\\
\\
\ddot{T}_{a}(t)=-\frac{A_{1}}{\tau_{1}^{2}}e^{-t/\tau_{1}}-\frac{A_{2}}{\tau_{2}^{2}}e^{-t/\tau_{2}}\\
\end{array} \right.
\end{equation}
For $t_{0}=0$ this leads to:
\begin{displaymath}
\begin{array}{l}
 T_{a}(0)=T_{0}=T_{1}-(A_{1}+A_{2})\\
\\
\Delta T=T_{1}-T_{0}=A_{1}+A_{2}
\end{array}
\end{displaymath}
Inserting solution (\ref{eq:solution-0}) into (\ref{eq:main-equation}) we obtain:
%
%\begin{widetext}
\begin{eqnarray*}
 -\frac{A_{1}}{\tau_{1}^{2}}e^{-t/\tau_{1}}-\frac{A_{2}}{\tau_{2}^{2}}e^{-t/\tau_{2}}-\left [\frac{k_{1}+k_{2}}{C_{a}}+\frac{k_{2}}{C_{p}}\right ]\left (\frac{A_{1}}{\tau_{1}}e^{-t/\tau_{1}}+\frac{A_{2}}{\tau_{2}}e^{-t/\tau_{2}}\right ) \\
\\
+\frac{k_{1}k_{2}}{C_{p}C_{a}}\left (\Delta T -A_{1}e^{-t/\tau_{1}}-A_{2}e^{-t/\tau_{2}}\right )=\frac{k_{2}P_{0}}{C_{p}C_{a}}\\
\end{eqnarray*}
%\end{widetext}
%
Isolating the two exponential terms the equation is fulfilled if
\begin{equation}
\label{eq:solution-1}
 \left\{ \begin{array}{lr}
e^{-t/\tau_{1}}\left [\frac{A_{1}}{\tau_{1}}\left (\frac{k_{2}}{C_{p}}+\frac{k_{1}+k_{2}}{C_{a}}\right )-\frac{A_{1}}{\tau_{1}^{2}}-\frac{A_{1}k_{1}k_{2}}{C_{p}C_{a}}\right ]=0 & \rm (i)\\
\\
e^{-t/\tau_{2}}\left [\frac{A_{2}}{\tau_{2}}\left (\frac{k_{2}}{C_{p}}+\frac{k_{1}+k_{2}}{C_{a}}\right )-\frac{A_{2}}{\tau_{2}^{2}}-\frac{A_{2}k_{1}k_{2}}{C_{p}C_{a}}\right ]=0 & \rm (ii)\\
\\
\frac{k_{1}k_{2}}{C_{p}C_{a}}(T_{1}-T_{0})-\frac{k_{2}P_{0}}{C_{p}C_{a}}=0 & \rm (iii)
\end{array} \right.
\nonumber
\end{equation}
From the \rm (iii) we obtain:
\begin{equation}\label{eq:power}
 P_{0}=k_{1}(T_{1}-T_{0})=k_{1}\Delta T
\end{equation}
and from \rm (i) and \rm (ii) we have the system of equations
\begin{equation}
\label{eq:solution-2}
 \left\{ \begin{array}{lr}
\frac{k_{2}}{C_{p}}+\frac{k_{1}+k_{2}}{C_{a}}=\frac{1}{\tau_{1}}+\frac{\tau_{1}k_{1}k_{2}}{C_{p}C_{a}}\\
\\
\frac{k_{2}}{C_{p}}+\frac{k_{1}+k_{2}}{C_{a}}=\frac{1}{\tau_{2}}+\frac{\tau_{2}k_{1}k_{2}}{C_{p}C_{a}}
\end{array} \right.
\nonumber
\end{equation}
with solutions
\begin{displaymath}
 \frac{1}{\tau_{1}}+\frac{\tau_{1}k_{1}k_{2}}{C_{p}C_{a}}=\frac{1}{\tau_{2}}+\frac{\tau_{2}k_{1}k_{2}}{C_{p}C_{a}}
\end{displaymath}
\begin{equation}\label{eq:final-1}
 C_{p}C_{a}=k_{1}k_{2}\tau_{1}\tau_{2}
\end{equation}
and (substituting $k_{2}$ into the first equation of the system):
\begin{displaymath}
 \frac{C_{a}}{k_{1}\tau_{1}\tau_{2}}+\frac{C_{p}}{k_{1}\tau_{1}\tau_{2}}+\frac{k_{1}}{C_{a}}=\frac{1}{\tau_{2}}+\frac{1}{\tau_{1}}
\end{displaymath}
\begin{equation}\label{eq:final-2}
 C_{p}=k_{1}\tau_{1}\left (1-\frac{k_{1}\tau_{2}}{C_{a}}\right ) +k_{1}\tau_{2}-C_{a}
\end{equation}
To obtain the $\tau_{2}$ constant we take the first of the equations (\ref{eq:heat-balance}) with the initial condition $t_{0}=0$ and $P(0)=P_{0}$ 
\begin{displaymath}
 P_{0}=C_{a}\dot{T}_{a}(0)+k_{1}[T_{a}(0)-T_{0}]+k_{2}[T_{a}(0)-T_{p}(0)]~,
\end{displaymath} 
where $T_{a}(0)=T_{p}(0)=T_{0}$. Considering Eq.~(\ref{eq:power}) we have:
\begin{displaymath}
 k_{1}(A_{1}+A_{2})=C_{a}\left (\frac{A_{1}}{\tau_{1}}+\frac{A_{2}}{\tau_{2}}\right )
\end{displaymath}
and therefore:
\begin{equation}\label{eq:tau2-0}
\tau_{2}=\frac{A_{2}\tau_{1}C_{a}}{(A_{1}+A_{2})\tau_{1}k_{1}-A_{1}C_{a}} 
\end{equation}
Considering now the simple case when $\tau_{2}=0$ (with, of course, $A_{2}=0$), we deduce:
\begin{equation}
\label{eq:solution-3}
 \left\{ \begin{array}{l}
T_{a}(t)=T_{1}-A_{1}e^{-t/\tau_{1}}\\
\\
C_{p}=k_{1}\tau_{1}-C_{a}\\
\\
P_{0}=k_{1}(T_{1}-T_{0})=k_{1}A_{1}\\
\end{array} \right.
\end{equation}
\vspace{1cm}

\end{document}